\RequirePackage{snapshot}

\documentclass[a4paper,twoside]{cernrep}

\makeatletter
\let\NAT@parse\undefined
\makeatother

\usepackage[numbers,sort,sectionbib]{natbib}
\usepackage[sectionbib]{bibunits}

\usepackage{times} 
\usepackage[T1]{fontenc} 
\usepackage[latin9]{inputenc} 
\usepackage{epic,eepic,pspicture,mathptmx,graphpap,epsf,titleref}

\usepackage{morefloats}
\usepackage{upgreek}
\usepackage{amsbsy,amsmath,amssymb} 
\usepackage{color,ulem} 
\usepackage{subfig}                 
\usepackage[printonlyused]{acronym} 
\usepackage[toc,page]{appendix}     
\usepackage{svn}                    
\usepackage{longtable}              
\usepackage{enumerate}		    
\usepackage{eurosym} 
\usepackage{multirow} 
\usepackage[loose]{units} 
\usepackage{threeparttable} 
\usepackage[point]{rccol} 
\usepackage{booktabs}
\usepackage{colortbl}
\usepackage{fancyhdr}
\usepackage{lineno}  
\usepackage{textpos} 
\usepackage{enumitem}
\setlist{noitemsep}
\setlist[itemize]{leftmargin=*, topsep=-0.3cm, partopsep=0pt, itemsep=0pt}
\setlist[enumerate]{leftmargin=*, topsep=0pt, partopsep=0pt, itemsep=0pt}

\usepackage[
     pdftex,
     plainpages        = false,
     bookmarks         = true,
     bookmarksnumbered = true,
     bookmarksopen     = true,
     pdfstartview      = FitH,
     pdfpagelayout     = SinglePage,
     colorlinks        = true,
     citecolor         = blue,
     linkcolor         = blue,
     urlcolor          = blue, 
     pdftitle={CLIC e+e- Linear Collider Studies},
     pdfauthor={CLIC},
     pdfkeywords={CLIC, accelerator, physics performances, detector concepts},
     pdfcreator={latex},
     pdfproducer={pdfLaTex}]{hyperref}

\def\roots {\ensuremath{\sqrt{s}}\xspace}

\def\epem       {\ensuremath{e^+e^-}\xspace}

\def\sqsq #1 {\ensuremath{\PSQ_{\mathrm{#1}}\PSQ_{\mathrm{#1}}}\xspace} 

\newcommand{\micron}{\ensuremath{\upmu\mathrm{m}}\xspace}

\newtoks{\abstract}{}

\begin{document}




 \begin{textblock*}{0mm}(135mm,-18mm)%
 \includegraphics[width=3cm]{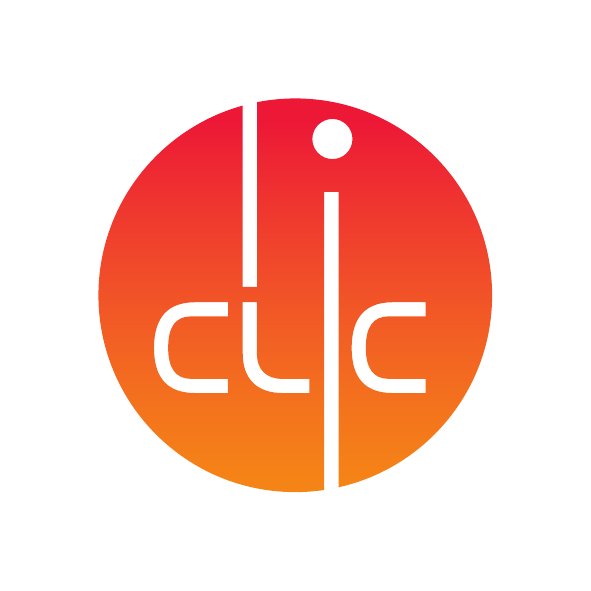}%
 \end{textblock*}%
\vspace{-2cm}
\begin{center}
{\huge  {CLIC} \epem Linear Collider Studies}\\\vspace{5mm}
{\LARGE  Input to the Snowmass process 2013}
\end{center}


\begin{center}
May 24, 2013
\end{center}

\noindent
This paper addresses the issues in question for Energy Frontier Lepton and Gamma Colliders by the Frontier Capabilities group of the Snowmass 2013 process and is structured accordingly. It will be accompanied by a paper describing the Detector and Physics studies for the CLIC project~\cite{c:snowmasspd} currently in preparation for submission to the Energy Frontier group.    

\begin{center}
Corresponding editors: Dominik Dannheim, Philippe Lebrun, Lucie Linssen,\\ Daniel Schulte, Steinar Stapnes.
\end{center}

\section{Introduction}

The Compact Linear Collider (CLIC) is a TeV scale high-luminosity linear \epem collider under development.
It is based on a novel two-beam acceleration technique providing acceleration gradients at the
level of 100 MV/m. Recent implementation studies for CLIC have converged towards a staged approach
offering a unique physics programme spanning several decades. In this scheme CLIC would provide 
high-luminosity  \epem collisions from a few hundred GeV to 3 TeV. 
The first stage, at or above the ~350 GeV top pair production threshold, 
gives access to precision Higgs physics through the Higgsstrahlung and WW 
fusion production processes,  providing absolute values of Higgs couplings to
both fermions and bosons. This stage also addresses precision top physics. 
The second stage, around 1.5 TeV, opens the energy frontier, allowing for the 
discovery of new physics phenomena. This stage also gives access to additional 
Higgs properties, such as the top-Yukawa coupling, the Higgs potential and rare 
Higgs decay branching ratios. The ultimate CLIC energy of 3 TeV enlarges the 
CLIC physics potential even further covering the complete scope for precision 
Standard Model physics, direct searches for pair-production of new particles with 
masses up to 1.5 TeV and optimal sensitivity to new physics and much higher 
masses through precision measurements. 
A staged implementation of CLIC along the lines described would open the door to an impressive long-term and timely physics programme at the
energy frontier, beyond the LHC programme.
This machine is therefore considered an important option for a post-LHC facility at CERN.

The feasibility studies for the CLIC accelerator have over the last
years systematically and successfully addressed the main technical
challenges of the accelerator project. Similarly, detailed detector
and physics studies confirm the ability to perform high-precision
measurements at CLIC.  

For more detailed descriptions we refer to the following
documents: \vspace{2mm}
\begin{itemize}
\item CLIC \epem Linear Collider Studies, eds. D.~Dannheim et al., 
submitted to the update process of the European Strategy for Particle Physics, 
July 2012~\cite{CLIC_Strategy_Input}; 
\item The Physics Case for an \epem Linear Collider, eds. J. Brau et
  al., submitted to the update process of the European Strategy for
  Particle Physics, July 2012~\cite{LC_Phys_Case_Strategy_Input};
\item A Multi-TeV Linear Collider based on CLIC Technology, CLIC
  Conceptual Design Report, 2012, eds.  M. Aicheler et
  al.~\cite{CLICCDR_vol1};
\item Physics and Detectors at CLIC, CLIC Conceptual Design Report,
  eds. L. Linssen et al.~\cite{CLICCDR_vol2};
\item The CLIC Programme: towards a staged \epem Linear Collider
  exploring the Terascale, CLIC Conceptual Design Report, 2012,
  eds. P. Lebrun et al.~\cite{CLICCDR_vol3}.
\end{itemize}
\vspace{2mm} The above CLIC CDR reports are supported by more than
1300
signatories\footnote{\href{https://edms.cern.ch/document/1183227/}{https://edms.cern.ch/document/1183227/}}
from the world-wide particle physics community.


\section{CLIC parameters and layout for a 3-stage implementation}
\begin{figure}[b!]
  \centering
  \includegraphics[scale=0.82]{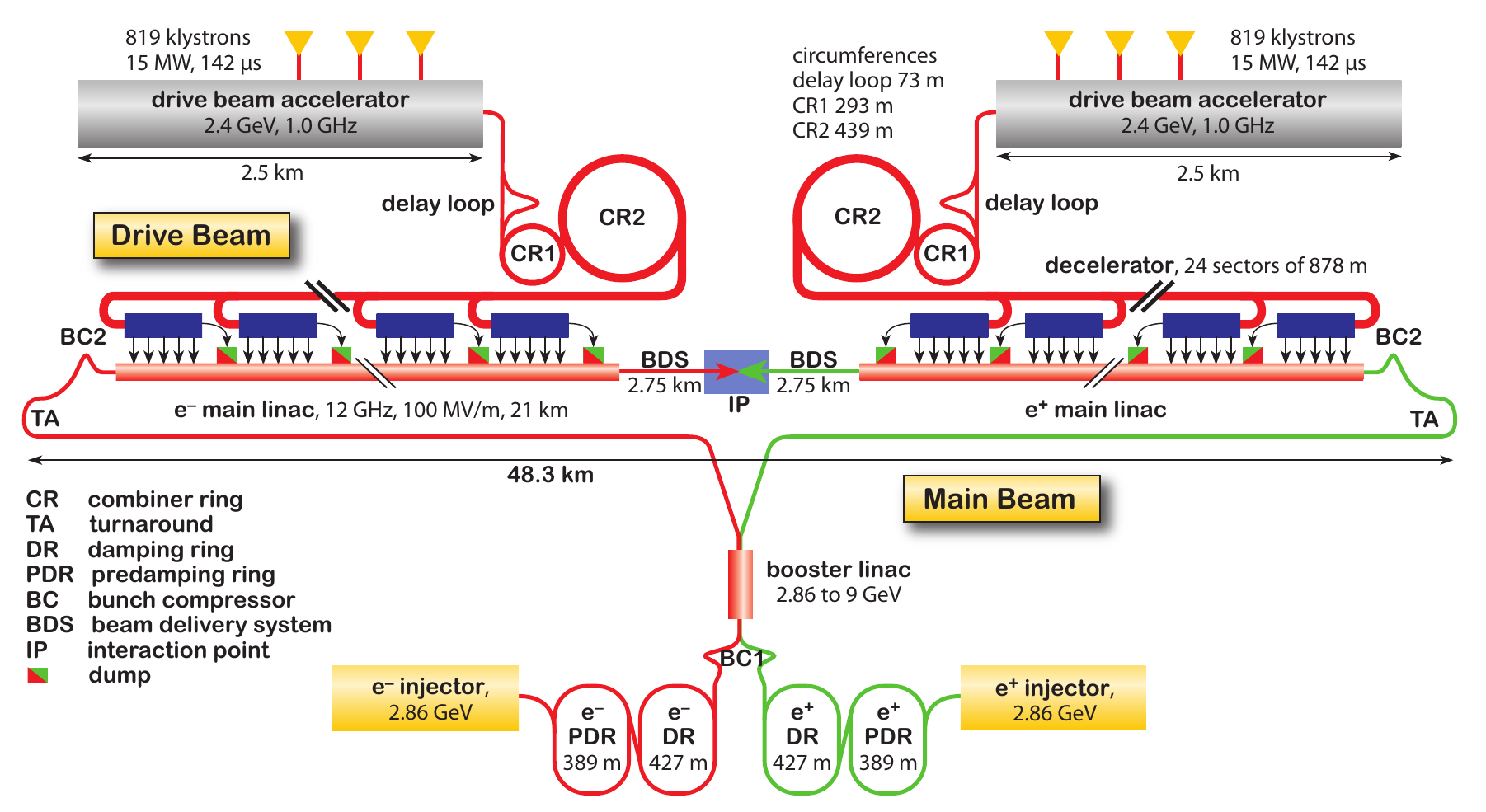}
  \caption{Overview of the CLIC layout at $\sqrt{s}=3$~TeV. The machine requires only one drive beam complex for stages 1 and 2.}
  \label{f:clic3tev}
\end{figure}

The CLIC layout at 3~TeV is shown in
Figure \ref{f:clic3tev}, and the
key parameters are given in Tables~\ref{t:1} and \ref{t:2}.  The
conceptual design is detailed in \cite{CLICCDR_vol1} and
\cite{CLICCDR_vol3}. 
The CLIC accelerator can be built in energy stages, re-using 
the existing equipment for each new stage.  At each energy stage the
centre-of-mass energy can be tuned to lower values within a range of a
factor three and with limited loss on luminosity performance.  Two
example scenarios of energy staging are given in~\cite{CLICCDR_vol3}
with stages of $500\;\rm GeV$, $1.4~(1.5)\;\rm TeV$ and $3\;\rm TeV$,
see Table~\ref{t:1} for scenario A and Table~\ref{t:2} for scenario B.
In both scenarios the first and second stage use only a single
drive-beam generation complex to feed both linacs, while in stage 3
each linac is fed by a separate complex. 

\begin{table}[ht!]
  \centering
  \caption{Parameters for the CLIC energy stages of scenario A.}
  \label{t:1}
  \begin{tabular}{l l l l l l}
    \toprule
    \textbf{Parameter} & \textbf{Symbol} & \textbf{Unit}& \textbf{Stage 1} & \textbf{Stage
      2} & \textbf{Stage 3}\\
    \midrule
    Centre-of-mass energy            & $\roots$              &GeV                                            & 500 & 1400 & 3000\\
    Repetition frequency             & $f_{rep}$             &Hz                                             & 50 & 50 & 50\\
    Number of bunches per train      & $n_{b}$               &                                               & 354 & 312 & 312\\
    Bunch separation                 & $\Delta\,t$           &ns                                             & 0.5 & 0.5 & 0.5\\
    \midrule
    Accelerating gradient       & $G$                        &MV/m                                           & 80 & 80/100 & 100\\
    \midrule
    Total luminosity                 & $\mathcal{L}$         &$10^{34}\;\textrm{cm}^{-2}\textrm{s}^{-1}$  & 2.3 & 3.2 & 5.9 \\
    Luminosity above 99\% of $\roots$& $\mathcal{L}_{0.01}$  &$10^{34}\;\textrm{cm}^{-2}\textrm{s}^{-1}$  & 1.4 & 1.3 & 2\\
    \midrule
    Main tunnel length          &                            &km                                             & 13.2 & 27.2 & 48.3\\
    Charge per bunch            & $N$                        &$10^9$                                         & 6.8 & 3.7 & 3.7\\
    Bunch length                & $\sigma_z$                 &$\micron$                                      & 72 & 44 & 44\\
    IP beam size                & $\sigma_x/\sigma_y$        &nm                                             & 200/2.6 & $\sim$ 60/1.5 &$\sim$ 40/1\\
    Normalized emittance (end of linac) & $\epsilon_x/\epsilon_y$    &nm                                             & 2350/20 & 660/20 & 660/20\\
    Normalized emittance (IP)        & $\epsilon_x/\epsilon_y$    &nm                                             & 2400/25 & --- & ---\\
    Estimated power consumption & $P_{wall}$                 &MW                                             & 272 & 364 & 589\\
    \bottomrule
  \end{tabular}
\end{table}
\begin{table}[ht!]
  \caption{Parameters for the CLIC energy stages of scenario B.}
  \label{t:2}
  \centering
  \begin{tabular}{l l l l l l}
    \toprule
    \textbf{Parameter}           & \textbf{Symbol} & \textbf{Unit}& \textbf{Stage 1} & \textbf{Stage
      2} & \textbf{Stage 3} \\
    \midrule
    Centre-of-mass energy             & $\roots$                &GeV                                           & 500 & 1500 & 3000\\
    Repetition frequency              & $f_{rep}$               &Hz                                            & 50 & 50 & 50\\
    Number of bunches per train       & $n_{b}$                 &                                              & 312 & 312 & 312\\
    Bunch separation                  & $\Delta\,t$            &ns                                            & 0.5 & 0.5 & 0.5\\
    \midrule
    Accelerating gradient                     & $G$                     &MV/m                                          & 100 & 100 & 100\\
    \midrule
    Total luminosity                  & $\mathcal{L}$           &$10^{34}\;\textrm{cm}^{-2}\textrm{s}^{-1}$ & 1.3 & 3.7 & 5.9 \\
    Luminosity above 99\% of $\roots$ & $\mathcal{L}_{0.01}$    &$10^{34}\;\textrm{cm}^{-2}\textrm{s}^{-1}$ & 0.7 & 1.4 & 2\\
    \midrule
    Main tunnel length                &                         &km                                            & 11.4 & 27.2 & 48.3\\
    Charge per bunch                  & $N$                     &$10^9$                                      & 3.7 & 3.7 & 3.7\\
    Bunch length                      & $\sigma_z$              &$\micron$                                    & 44 & 44 & 44\\
    IP beam size                      & $\sigma_x/\sigma_y$     &nm                                            & 100/2.6 & $\sim$ 60/1.5 & $\sim$ 40/1\\
    Normalized emittance (end of linac)    & $\epsilon_x/\epsilon_y$ &nm                                            & --- & 660/20 & 660/20\\
    Normalized emittance                   & $\epsilon_x/\epsilon_y$ &nm                                            & 660/25 & --- &---\\
    Estimated power consumption       & $P_{wall}$              &MW                                            & 235 & 364 & 589\\
    \bottomrule
  \end{tabular}
\end{table}

Staging scenario A aims at achieving high luminosity at 500~GeV
collision energy with increased beam current. This requires
larger apertures in the accelerating structures which
therefore operate at a lower gradient. The re-use of these structures
in the second stage limits the achievable collision energy to
1.4~TeV. Staging scenario B aims at reducing the cost of the 500~GeV
stage using full-gradient accelerating structures at nominal beam
current, resulting in lower instantaneous luminosity.  The re-use of
these structures allows reaching 1.5~TeV collision energy in the
second stage.

The recent LHC Higgs discovery makes an initial energy stage at 375\;GeV instead of 500\;GeV attractive, but final choices depend on future LHC findings.
While the CDR design has been optimized for the 3\:TeV stage only, the accelerator design is now being re-optimized also at the initial stages and
using the improved understanding of the cost and power consumption obtained during the preparation of the CDR.
In case of growing interest in a lower energy Higgs
factory, studies of a klystron-based initial stage with a faster
implementation could become part of this evaluation. 

\section{Main technical challenges and demonstrators}
The CLIC design is based on three key technologies, which have been
addressed experimentally:\vspace{2mm}
\begin{itemize}
\item The use of normal-conducting accelerating structures in the main
  linac with a gradient of $100\;\rm MV/m$, in order to limit the
  length of the machine.  The RF frequency of $12\;\rm GHz$ and
  detailed parameters of the structure have been derived from an
  overall cost optimisation at 3~TeV. Experiments at KEK, SLAC and
  CERN verified the structure design and established its gradient and
  breakdown-rate performance.
\item The use of drive beams that run parallel to the colliding beams
  through a sequence of power extraction and transfer structures,
  where they produce the short, high-power RF pulses that are
  transferred into the accelerating structures. These drive beams are
  generated in a central complex.  The drive-beam generation and use
  has been demonstrated in a dedicated test facility (CTF3) that has
  been constructed and operated for many years at CERN by the
  CLIC/CTF3 collaboration.
\item The high luminosity that is achieved by the very small beam
  emittances, which are generated in the damping rings and maintained
  during the transport to the collision point. These emittances are
  ensured by appropriate design of the beam lines and tuning
  techniques, as well as by a precision pre-alignment system and an
  active stabilisation system that decouples the magnets from the
  ground motion. Prototypes of both systems have demonstrated
  performance close to or better than the specifications.
\end{itemize} \vspace{2mm} 
Related system parameters have been benchmarked in CTF3, in advanced light
sources, in the ATF(2) and CesrTA, and in other setups. 
In addition, a broad technical development programme has successfully
addressed many critical components. Among them are those of the main
linac, which are most important for the cost, and their integration
into modules. The drive-beam components have largely been addressed in
CTF3. Other performance-critical components have been developed and
tested, e.g., the final focus magnets, which will be located in the
detector and need to provide a very high field, and high-field damping
ring wigglers, which rapidly reduce the beam emittances.  Design
studies foresee 80\% polarisation of the electrons at collision, and
the layout is compatible with addition of a polarized positron source.
The successful validation of the key technologies and of the critical
components establish confidence that the CLIC performance goals can be
met.

\section{Machine footprint, power and cost}
\begin{figure}[tb]
  \centering
  \includegraphics[width=130mm]{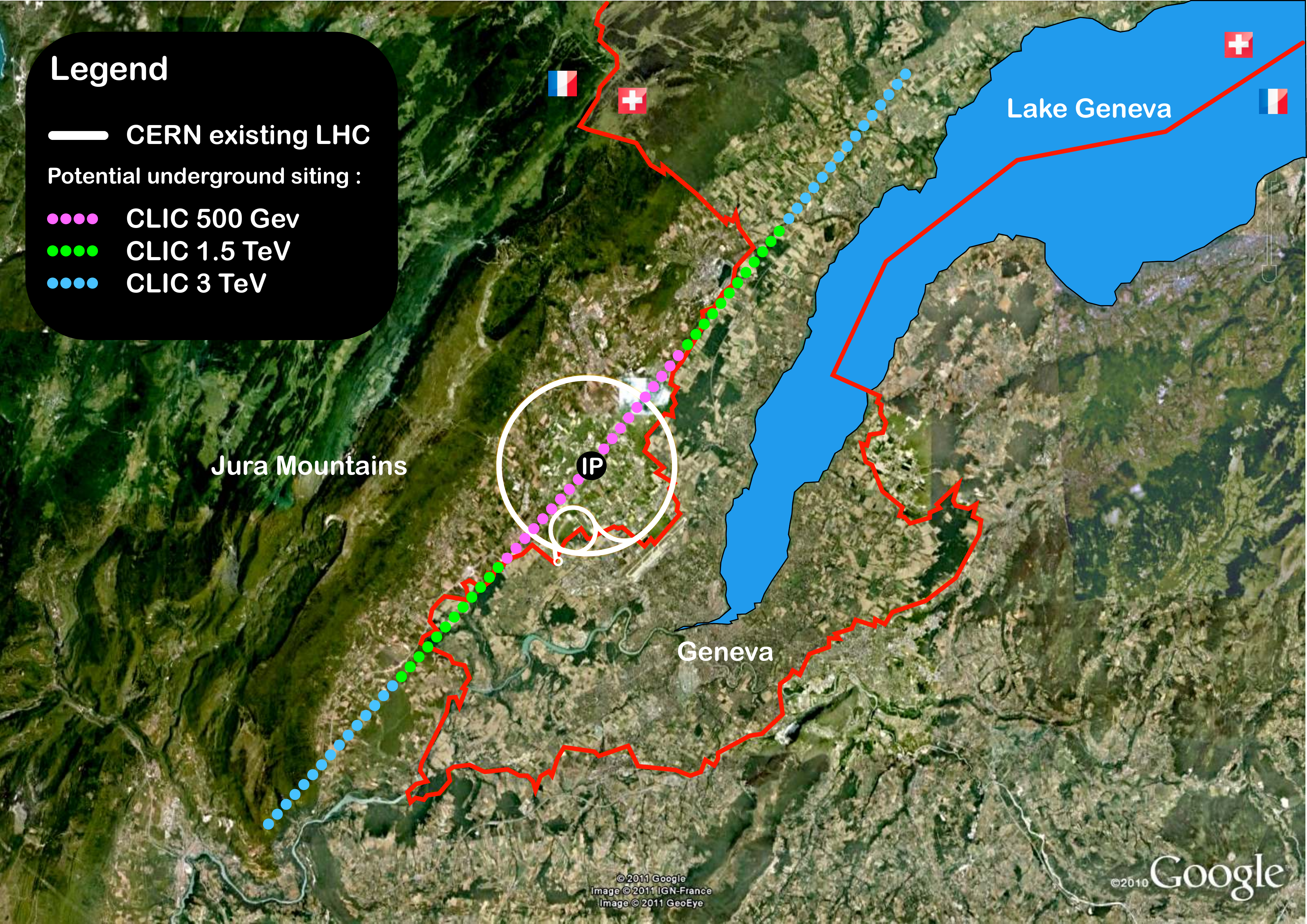}
  \caption{CLIC footprints near CERN, showing various implementation
    stages. The main tunnel lengths are given in given in Tables~\ref{t:1} and \ref{t:2}.}
  \label{fig:site}
\end{figure}

Detailed site studies show that CLIC can be implemented underground
near CERN, with the central main and drive beam complex on the CERN
domain, as shown in Figure~\ref{fig:site}. The site specifications do
not constrain the implementation to this location.

The current CLIC parameters are the result of a global cost optimisation at
$3\;\rm TeV$, see Chapter 2.1 in~\cite{CLICCDR_vol1}.  However, the technology
can be used effectively over a wide range of centre-of-mass
energies as explained above, and the design re-optimized for different objectives if desired.

The nominal electrical power consumption of all accelerator systems
and services, including the experimental area and the detectors and
taking into account network losses for transformation and distribution
on site, is given in Table~\ref{tab:power} for staging scenarios A and
B.  The table also shows residual power consumption without beams for
two modes corresponding to short ("waiting for beam") and long
("shutdown") beam interruptions. The large variations and volatility
of power consumption will allow CLIC to be operated as a peak-shaving
facility, matching the daily and seasonal fluctuations in power demand
on the network.  Several paths aiming at reducing power consumption or
improving the energy footprint of the machine have been identified and
are under investigation, e.g., reduction of design current density in
magnet windings and cables, replacement of normal-conducting by
permanent or super-ferric magnets, development of high-efficiency
klystrons and modulators, recovery and valorisation of waste heat.
Furthermore, the on-going work to optimize the energy stages of CLIC will include power 
reduction as a key parameter.

\begin{figure}[ht!]
  \centering
  \includegraphics[width=0.7\textwidth]{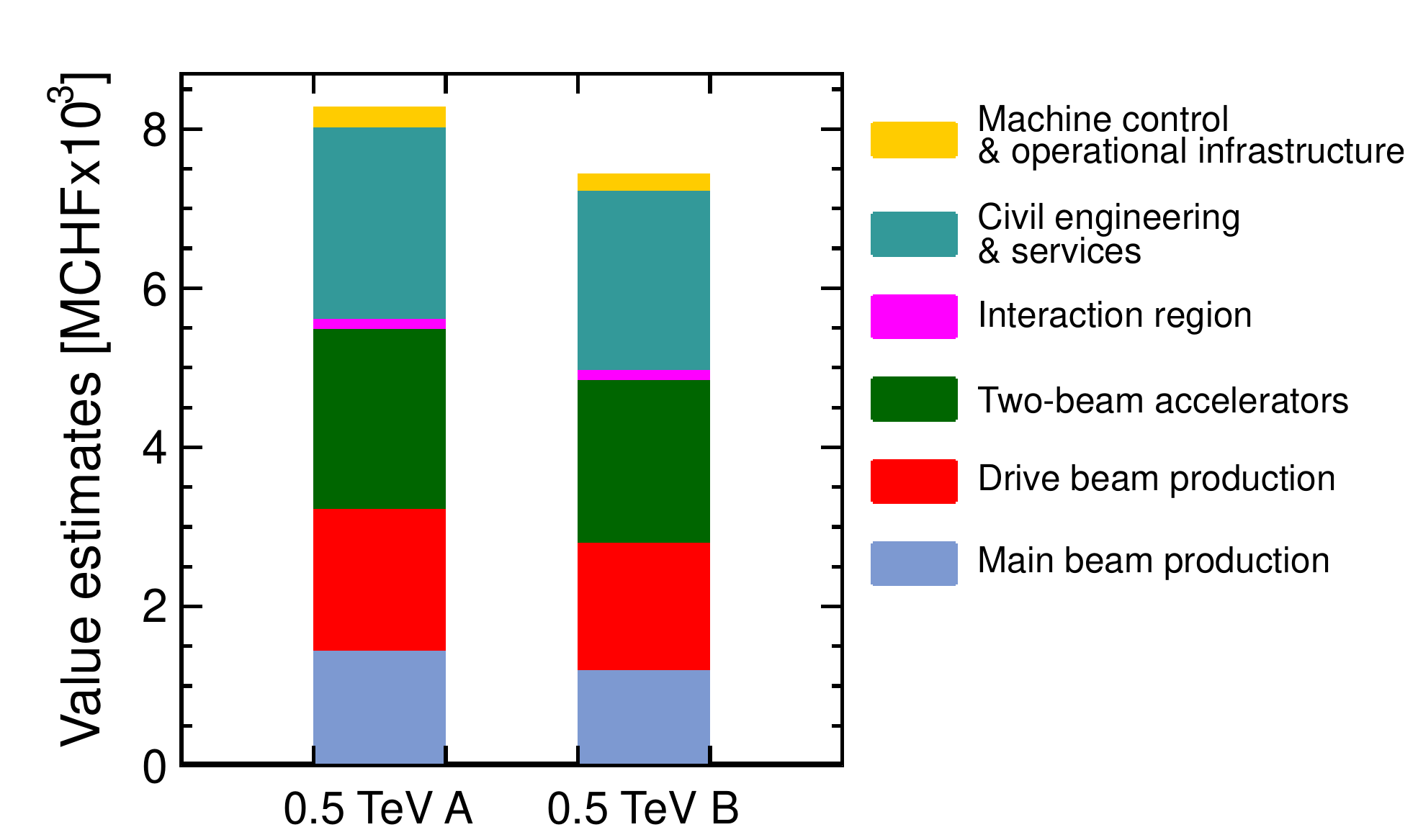}
  \caption{Cost structure of the CLIC accelerator complex at 500 GeV
    for scenarios A and B.}
  \label{fig:cost_500GeV}
\end{figure}

\begin{table}[ht!]
  \caption{\label{tab:power}CLIC power consumption for staging
    scenarios A and B.}
  \centering
  \begin{tabular}{c l l l l }
    \toprule
    Staging scenario &
    $\roots$ [TeV] &
    $P_{\textrm{nominal}}[\textrm{MW}]$ &
    $P_{\textrm{waiting for beam}}[\textrm{MW}]$ &
    $P_{\textrm{shutdown}}[\textrm{MW}]$ \\
    \midrule \midrule
    &  0.5 & 272 & 168 & 37\\
    A & 1.4 & 364 & 190 & 42\\
    &  3.0 & 589 & 268 & 58\\
    \midrule
    &  0.5 & 235 & 167 & 35\\
    B &  1.5 & 364 & 190 & 42\\
    &  3.0 & 589 & 268 & 58\\
    \bottomrule
  \end{tabular}
\end{table}
\begin{table}[ht!]
  \caption{\label{tab:cost}Value and labour estimates of CLIC 500 GeV.}
  \centering
  \begin{tabular}{c l l }
    \toprule
    Staging scenario &
    Value [MCHF] &
    Labour [FTE years]\\
    \midrule \midrule
    A  & $8300^{+1900}_{-1400}$ & 15700 \\[4pt] 
    B  & $7400^{+1700}_{-1300}$ & 14100 \\
    \bottomrule
  \end{tabular}
\end{table}
The cost estimates
follow the ``value'' and ``explicit labour'' methodology used for the ILC
Reference Design report~\cite{ILC_RDR_vol3}.  They are based on the
work breakdown structures established for the different stages of the
two scenarios, and on unit costs obtained for other similar supplies
or scaled from them, and from specific industrial
studies. Uncertainties include technical and procurement risks, the
latter being estimated from a statistical analysis of procurement for
the LHC. The value estimates are expressed in Swiss francs (CHF) of
December 2010 and can thus be escalated using relevant Swiss official
indices. Explicit labour is estimated globally by scaling from LHC
experience. The results are given in Table~\ref{tab:cost}. The cost
structure of the accelerators at 500~GeV collision energy for staging
scenarios A and B is illustrated in Figure~\ref{fig:cost_500GeV}. The
incremental value from the first to the second stage is about 4
MCHF/GeV (scenario B).  Potential savings have been identified for a
number of components and technical systems, amounting to about 10\% of
the total value.  Examples of such savings are the substitution of the
hexapods for the stabilisation of the main-beam quadrupoles with beam
steering, the doubling in length of the support girders for the
two-beam accelerator modules, or the alternative of using assembled
quadrants instead of stacked disks for construction of the
accelerating structures.  Moreover, significant additional savings are
expected from re-optimizing the design of the chosen
energy stages.

\section{Preparation timeline, project development and construction schedules}

\begin{figure}[t!]
  \centering
  \includegraphics*[width=\textwidth]{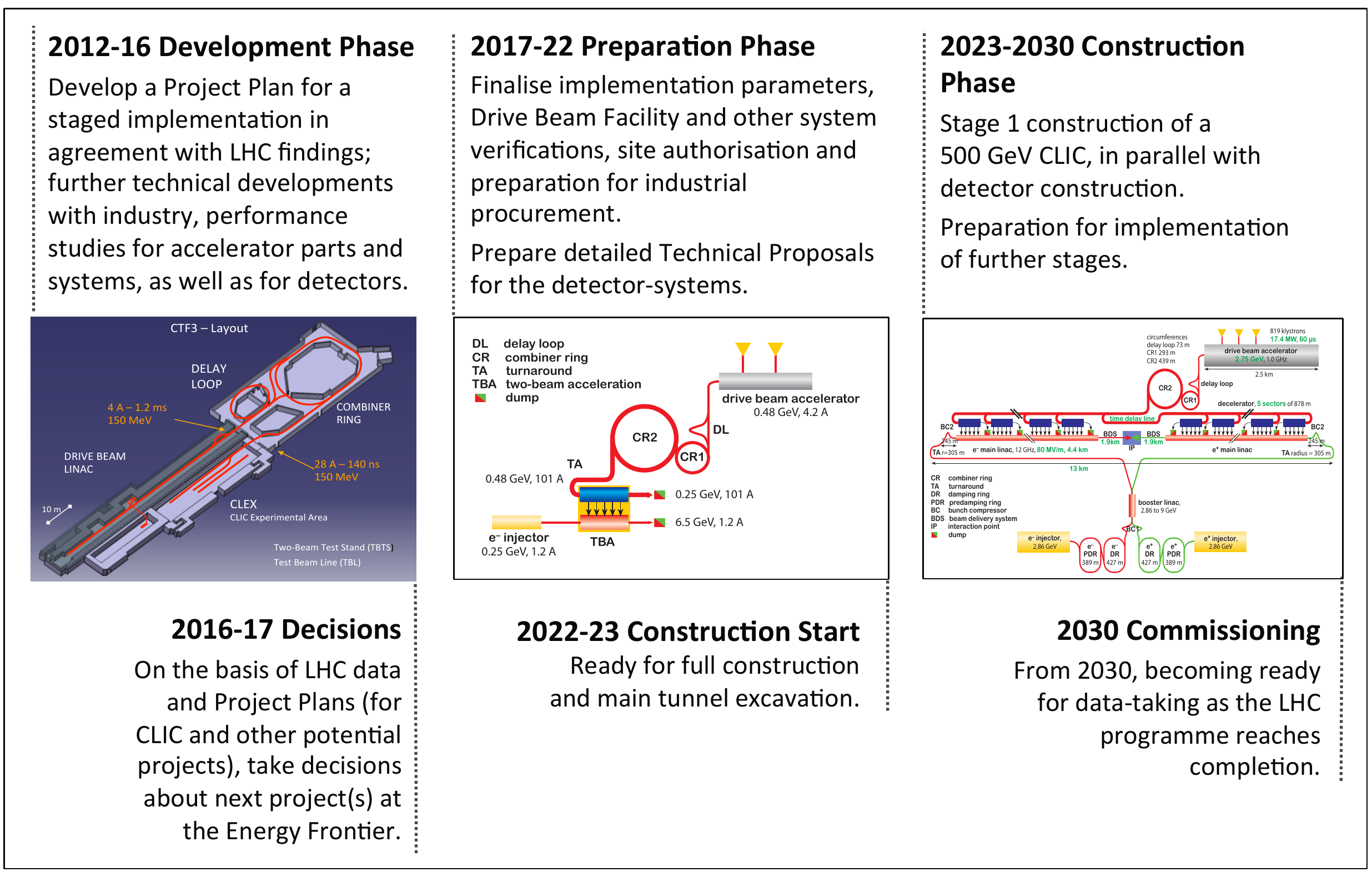}
  \caption{Top row: An outline of the CLIC project timeline with main
    activities leading up to and including the first stage
    construction. Middle row: illustrations of the CTF3 facility (one
    of several testing facilities of importance to the project
    development), a new large drive beam facility with final CLIC 
    elements
    which is also needed for acceptance tests, and a 500 GeV
    implementation.  Bottom row: Main decision points and activity
    changes.} 
  \label{f:timeline}
\end{figure}

A development programme for the {CLIC} project has been established and
is being carried out concurrently with {LHC} operation at 8~TeV and
later full energy, covering the period until 2016-17. By that time both
the {LHC} physics results and technical developments should have
reached a maturity that would allow a decision about the most
appropriate next project at the energy frontier beyond LHC. The major
contenders are a Linear Collider with multi-TeV capabilities or an energy-upgraded {LHC}.

The CLIC development programme will lay the ground work for a complete Project
Implementation Plan for the project, planned to be available by 2016-17.
One important element is that a re-baselining of the CLIC energy stages is underway as explained above.
Cost and power studies will follow; in addition to being key elements for the stage and design optimization additional technical developments can lead to important reductions.   
Important technical studies will address stability and alignment, timing and phasing, stray fields,
and dynamic vacuum including collective effects. Other studies will address failure modes and operation issues.  The collaboration will continue to identify and carry out system tests,
and priorities are the measurements in CTF3, ATF, and related to the CLIC injector. Further X-band structure development and tests are high priorities as well as constructing integrated modules where a number of central functional elements are included and need to be optimized.
Initial site studies have already been carried out and preliminary footprints have been identified for an initial machine as well as an ultimate 3 TeV layout, and these studies will continue. The 48 CLIC institutes are all participating in the planning and execution of these activities, and the programme adapted to the foreseen resources available in this period. Several new institutes have joined or are in the process of joining the studies. This programme will put the CLIC project in a position to be ready by 2017, i.e. after two years of LHC data-taking at full energy, for a decision on a future post-LHC facility at the energy frontier. 

Construction start for {CLIC} could be around 2023 after an initial Project Preparation Phase
2017--2022. During the Preparation Phase it is essential to optimise the component performances and to reduce their cost, in preparation of large industrialization contracts. In addition, a number of key system performances need to be addressed to minimise the risk of the CLIC project implementation. In this way requirements can be rationalized
by understanding the interplay between safety margins and therefore the overall project cost can be reduced. 
The drive beam and luminosity performances, in particular, are best addressed in larger system tests. 
The currently foreseen timeline for the
CLIC project preparation and stage 1 construction is shown in Figure~\ref{f:timeline}, with details
presented in ~\cite{CLICCDR_vol3}.

\begin{figure}[ht!]
  \centering
  \includegraphics[width=\textwidth]{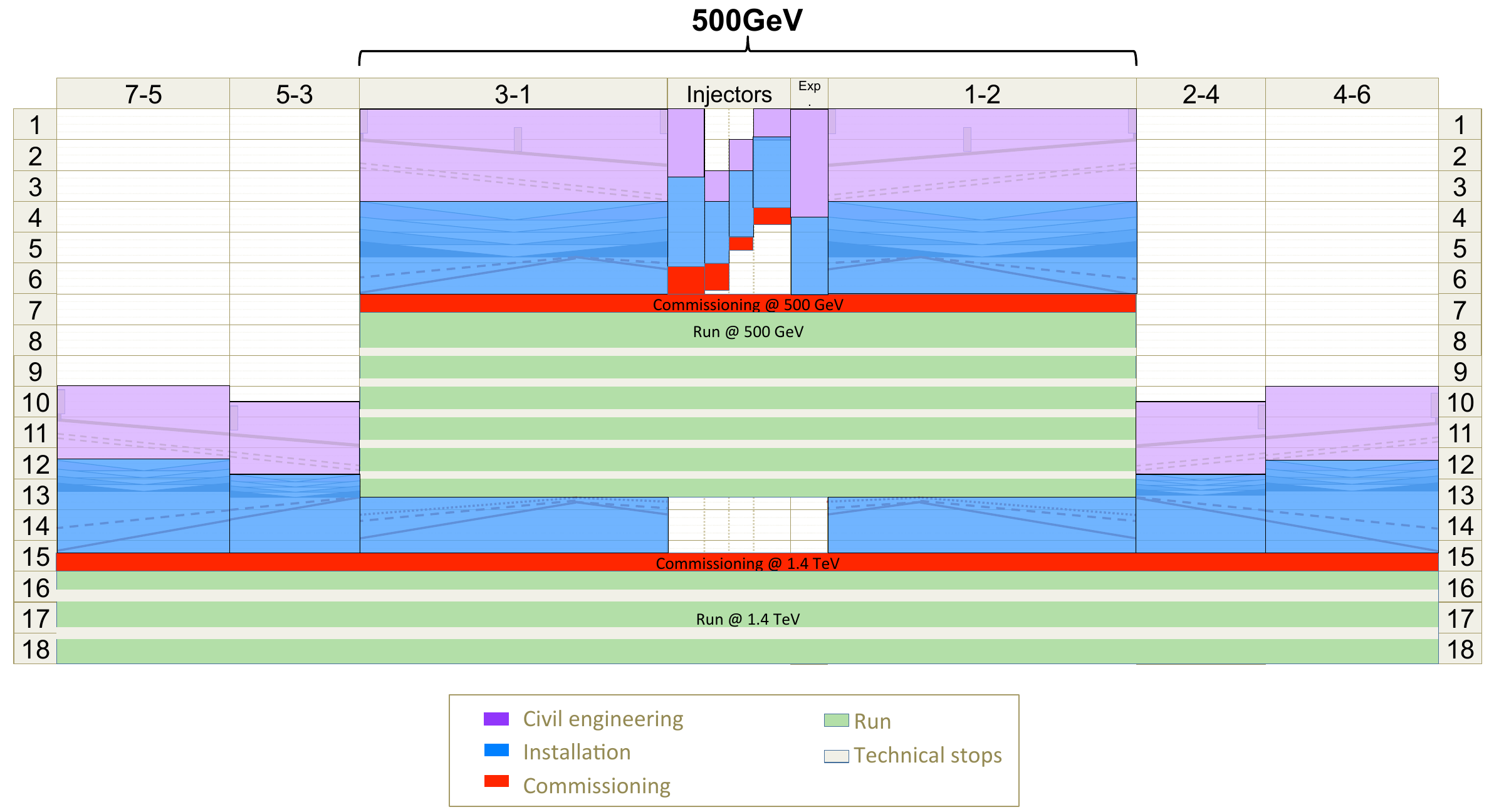}
  \caption[]{Overall ``railway'' schedule for the first two stages of
    scenario A.  The horizontal scale is proportional to tunnel
    length, with the experimental area in the centre. The vertical
    scale shows years from the start of construction.  The
    construction schedule for the main-beam, the drive-beam injectors
    and the experimental area are shown in the centre.}
  \label{fig:overallSchedule_scenarioA}
\end{figure}

Construction schedules (Figure~\ref{fig:overallSchedule_scenarioA})
are essentially driven by civil engineering, infrastructure and
machine installation.
Production of the large-series components proceeds at rates such that
they become available for installation as soon as preceding
construction activities allow it. In the first stage, construction of
the injector complex, experimental area and detectors just matches the
construction time for the main linacs, thus allowing commissioning
with beam to start in year~7. This would allow completion of the stage 1 project by 2030 when the 
LHC programme reaches a natural completion. 
In order to minimize interruption of operation for physics, civil
engineering and series component production for the second stage must
re-start in year~10, thus allowing commissioning in year~15 (scenario
A): this can be achieved without interference with operation for
physics in the first stage.

\section{Luminosity goals and possible overall project timeline for CLIC}

With the recent discovery of a new Higgs-like state at
$\sim 125$~GeV at {LHC} and considering the importance of studies
near the top threshold, it is evident that
an initial {CLIC} stage at 375~GeV will already provide
exceptional physics. 
A second stage around ~1.2--1.5~TeV would allow for
measurements of other Higgs properties, such as the top-Yukawa coupling,
the Higgs self-coupling and rare Higgs decay modes.
With the present knowledge a third stage well beyond
1.5~TeV can only be justified by the general arguments of improved
production cross-sections and precision on the measurements mentioned
above, and a significantly increased search capability.  It is however
important to keep in mind that the very recent results from {LHC} open
a completely new experimental territory. We can look forward to more
{LHC} results from the 2012 data-analyses and in particular when {LHC} moves to full energy
running in 2015, potentially providing even more exciting prospects
for a future {CLIC} programme, including ultimate energy stages
beyond ~1.5 TeV.

\begin{figure}[t!]
  \centering
  \includegraphics[width=0.45\textwidth]{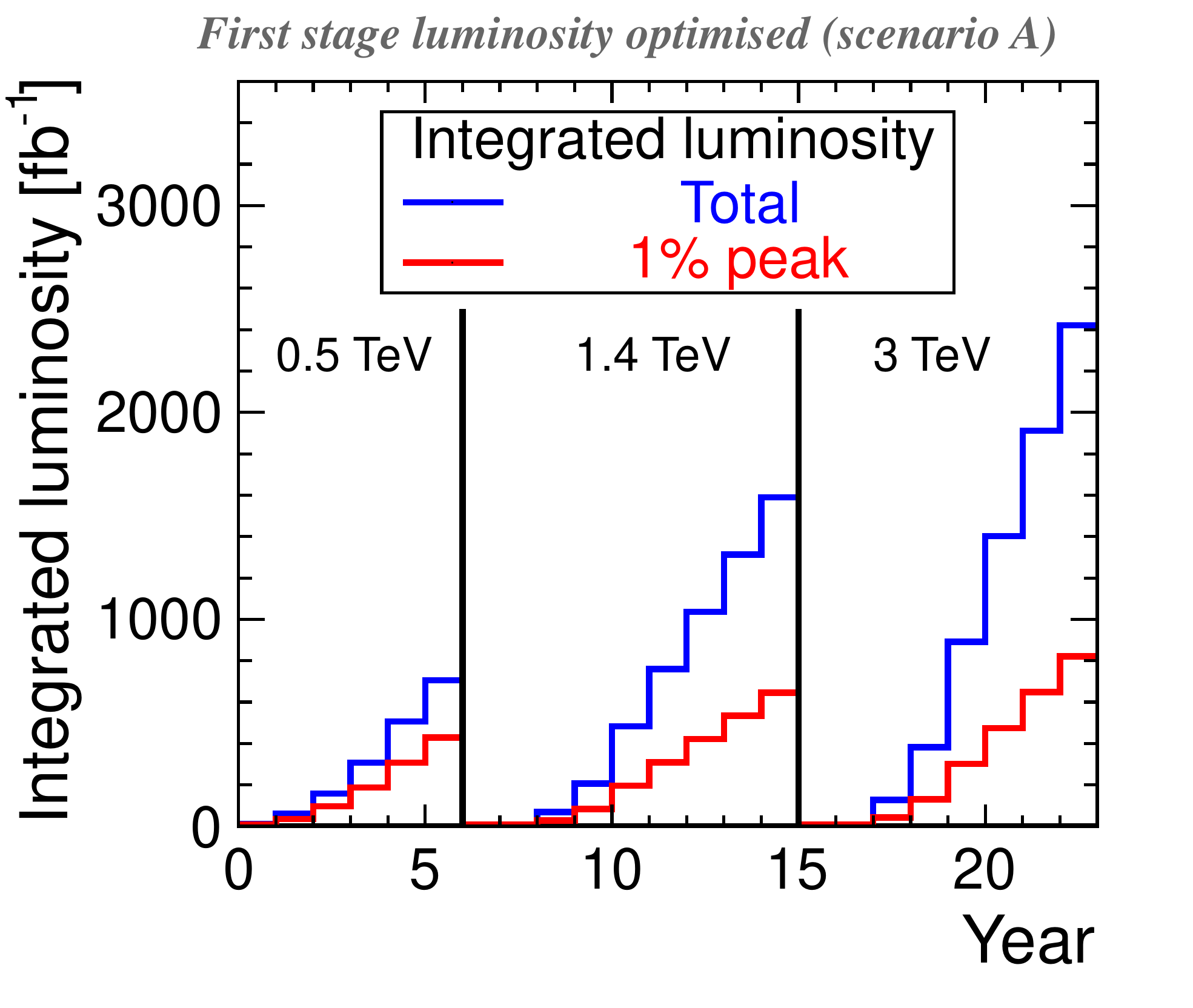}
  \includegraphics[width=0.45\textwidth]{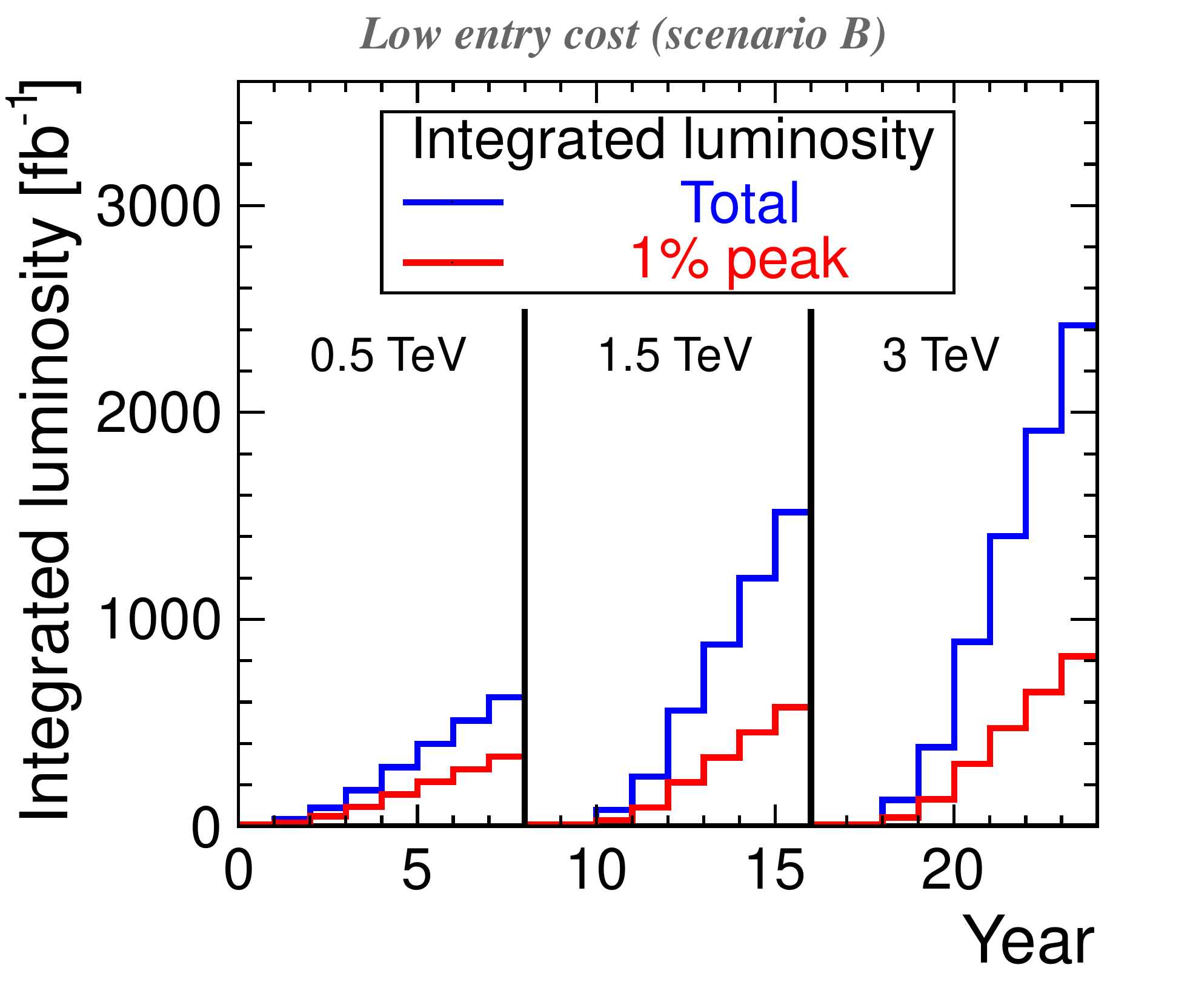}
  \caption{Integrated luminosity in the scenarios optimized for
    luminosity in the first energy stage (left) and optimized for
    entry costs (right). Years are counted from the start of beam
    commissioning.  These figures include luminosity ramp-up of four
    years (5\%, 25\%, 50\%, 75\%) in the first stage and two years
    (25\%, 50\%) in subsequent stages.}
  \label{fig:integratedLumi}
\end{figure}

The {CLIC} project as outlined is an ambitious long-term
programme, with an initial 7 year construction period and three energy
stages each lasting 6--8 years to achieve the integrated luminosity goals, interrupted by 2 year upgrade periods.
Possible operating scenarios for the complete CLIC programme
are sketched in Figure~\ref{fig:integratedLumi}: the duration of each
stage is defined by the integrated luminosity targets of 500 fb$^{-1}$
at 500 GeV, 1.5 ab$^{-1}$ at 1.4~(1.5)~TeV and 2~ab$^{-1}$ at~3 TeV
collision energy. The integrated luminosity in the first stage can be
obtained for scenario B by operating for two more years; this is
partly regained in the next stage, so that the overall duration of the
three-stage programme is comparable for both cases, about 24 years
from start of operation.

\section{Technology applications}
Several of the central technologies in the CLIC project have applications for and are being
developed with other communities. The most important examples of the use of high-gradient normal-conducting technology developed for CLIC are:
\begin{itemize}
\item The development of compact linacs for proton and carbon ion cancer treatment, e.g. the TULIP proposal
of the TERA project~\cite{c:TULIP}.
\item Future free electron lasers (FELs) for photon-science, which encompasses biology,
chemistry, material science and many other fields. Common developments have been made with the SWISS FEL~\cite{c:SWISSFEL} and the
ZFEL~\cite{c:ZFEL} and further collaborations with several other projects are being prepared.
\item Compton-scattering gamma ray sources providing MeV-range photons for laser-based nuclear physics (nuclear-photonics) and
fundamental processes (QED studies for example)~\cite{c:ELI}. 
There are also potential applications such as nuclear resonance fluorescence
for isotope detection in shipping containers and mining.
\end{itemize}
Also synchrotron-based light sources and the CLIC damping rings share similar issues and challenges, which are addressed in a collaborative effort.




\bibliography{Main}

\end{document}